\documentclass[10pt,preprint,letterpaper]{article}

\usepackage{amsfonts}
\usepackage{amsmath}
\usepackage{amssymb}
\usepackage{graphicx}
\usepackage[numbers,sort&compress]{natbib}
\title{High sensitivity ultraviolet graphene-metamaterial integrated electro-optic modulator enhanced by superlubricity}
\author{\small Xu Yan-li$^{1,2}$, Li Wei-min$^1$, Li Rong$^1$, Liu Jiang-tao$^{1*}$, Liu Ze$^{3\dag}$, Wu Zhen-Hua$^{2,4\ddag}$\\  \footnotesize
$^{1}$College of  Mechanical and electrical engineering, Guizhou Minzu University, Guiyang 550025, China \\
\footnotesize  $^{2}$Key Laboratory of Microelectronic Devices and Integrated Technology, Institute \\ \footnotesize of Microelectronics, Chinese Academy of Sciences, Beijing 100029, China\\ 
 \footnotesize $^{3}$
Department of Engineering Mechanics, School of Civil Engineering, \\ \footnotesize Wuhan University, 430072, Wuhan, China\\
 \footnotesize $^{4}$
University of Chinese Academy of Sciences, Beijing, 100029, China\\
 \footnotesize   $^{*}$Email: jtliu@semi.ac.cn
  \footnotesize   $^{\dag}$Email: ze.liu@whu.edu.cn
  \footnotesize  $^{\ddag}$ Email:wuzhenhua@ime.ac.cn
 }

\begin{document}

\maketitle

\begin{abstract}
Ultraviolet (UV) electro-optic modulation system based on graphene-plasmonic metamaterials nanomechanical system (NEMS) with superlubricity is investigated. Due to the strong optical absorption intensity of graphene in the UV region and the combination of metamaterial structure based on surface plasmons, the modulation depth of the UV NEMS electro-optic modulator approaches as high as 8.5 times compared to the counterpart modulator in visible light region. Meanwhile, the superlubricity significantly reduces the power consumption of the UV electro-optic modulation system due to its extremely low friction coefficient. It also significantly improves the response speed of the modulator, which can reach the order of a nanosecond. The modulation voltage can be lower than 200 mV. The proposed electro-optic modulation system has a simple structure and high sensitivity, which is supposed to have important applications in UV optoelectronic devices and systems.
\end{abstract}

\section*{Introduction}

Ultraviolet (UV) light is the electromagnetic wave whose wavelength is between 10 and 400 nm in the electromagnetic spectrum \cite{C:201806006,glaser1968power,averin2016wavelength}. It has important applications in spectral analysis, UV imaging, UV confidential optical communication, and other military and civil fields \cite{lin2019diamond,liao2015uv,li2019solar}. Since the working wavelength of photoelectric devices depends on the bandgap of semiconductor materials, UV optoelectronic devices are usually prepared using traditional wide bandgap semiconductors, such as GaN, SiC, ZnO, Ga$_2$O$_3$, and AlGaN \cite{razeghi1996semiconductor,goldberg1999semiconductor,alaie2015recent,rahman2019zinc}. However, UV modulation is mainly realized through internal modulation, with a low modulation speed and high noise. For example, the  modulation speed of UV modulation using mercury-xenon lamp as the light source is about 400 kHz \cite{puschell1990high,han2012theoretical,xu2008ultraviolet}. Graphene has a very strong light response in the  UV region due to the strong electron-hole interaction near the saddle point of graphene, thus graphene is an ideal UV photoelectric material \cite{yang2009excitonic,boosalis2012visible,cai2018ultraviolet,xu2020ultraviolet}. However, graphene in the UV region has a high Fermi energy level. Therefore, it is not easy to adjust the light absorption near the Fermi energy level using traditional modulation of graphene Fermi energy level \cite{guo2021polarization}.

Recently, Liu $et$ $al.$ introduced the high-speed spontaneous recovery motion based on superlubricity into the graphene micro-nano mechanical system, and designed a micro-nano mechanical electro-optic modulator with a high speed and low power consumption based on superlubricity \cite{zhou2021ultrawide}. This kind of modulator does not need to adjust the Fermi energy of graphene. Superlubricity is the interface with a very small coefficient of friction \cite{hod2018structural,zhai2019nanomaterials,liu2017robust,cihan2016structural,li2020microscale,berman2018approaches,zheng2002multiwalled,li2020toward,luo2021origin,gao2021computational}, which will lead to some new applications of micro-nano mechanical systems, such as high-speed self-recovery graphite blocks \cite{yang2013observation,liu2012interlayer}. The modulator uses the superlubrication between the interfaces to prevent wear, reduce friction energy loss, improve the sensitivity and working life of micro-nano mechanical systems, and reduce power consumption. This modulation is based on the high-speed, low-loss mechanical motion of graphene and does not require a high Fermi energy level. Therefore, we introduce a micro-nano mechanical system based on superlubrication into the UV electro-optic modulator. Meanwhile, to improve the performance of the device, the optical signal modulation is realized by combining the superlubricity with plasmonic metamaterials; thereby, improving the modulation depth of the electro-optic modulator.

The most significant feature of plasmonic metamaterials is that they can be directly coupled with incident light to stimulate the surface plasmon resonance, which can be tuned by changing the parameters of metal nanostructure, material, and external environment, with higher degree of freedom and flexibility \cite{schuller2010plasmonics,willets2007localized,hutter2004exploitation,leng2021enhanced}. Based on these, they utilized the advantage of the selective light absorption and scattering of plasmonic metamaterial structures, local electric field constraints, and other properties to enhance light-matter interactions. It has been widely used in light modulator \cite{liu2011all,wang2015coupling,babicheva2015transparent,vinnakota2020plasmonic}, light detector \cite{sobhani2013narrowband,miao2015surface}, photovoltaic solar energy \cite{atwater2011plasmonics,clavero2014plasmon}, biological refractive index sensor \cite{cao2014gold,unser2015localized}, surface plasmon laser \cite{noginov2009demonstration,ma2012multiplexed,ma2021plasmonic}, photochemical catalysis \cite{wang2013h,hou2013review}, surface enhanced Raman spectroscopy \cite{wei2013hot,ding2016nanostructure}, and optical holographic imaging \cite{montelongo2014plasmonic,huft2017holographic}. With the development of theoretical research, plasmonic metamaterials have more overlaps with biology, chemistry, energy, electronics, materials, and other disciplines. Moreover, their applications are more extensive.

Therefore, this study investigates a UV electro-optic modulation system based on a graphene-plasmonic metamaterial nanomaterial electromechanical system (NEMS) with superlubricity. Superlubricity further improves the performance of the UV photoelectric modulator based on graphene-plasmonic metamaterial NEMS due to its extremely low friction coefficient. The modulation voltage is lower than 200 mV, and the modulation response speed can be up to nanoseconds. Meanwhile, the modulation depth of the electro-optic modulator is further improved by combining with the strong local field formed by the plasmonic metamaterials in the UV region. The absolute variation of reflectance is more than 0.4. Additionally, the polarization-dependent characteristics of the modulator based on plasmonic metamaterials make it possible to develop novel polarization-dependent UV devices. The optical modulation system has a simple structure and high sensitivity, which will have important applications in UV communication.

\section*{Theoretical model and analytical method}

\begin{figure}
	\centering
	\includegraphics[width=\linewidth,clip]{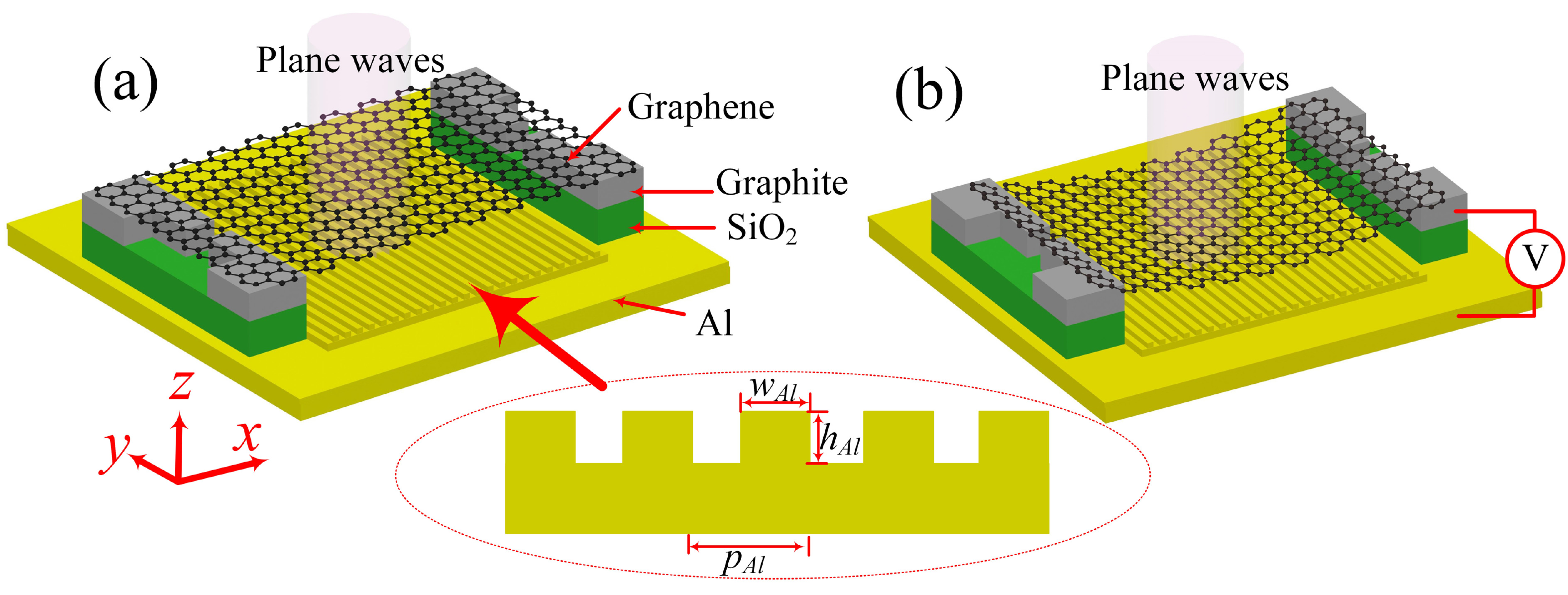}
	\caption{(a) Schematic of UV modulation structure; (b) Schematic of graphene moving downward and shrinking after applying voltage. Electrodes are connected to a graphite and plasmonic metamaterial substrate.}
	\label{Fig1}
\end{figure}

The structure studied is shown in Fig. \ref{Fig1}. Due to the high reflectivity of aluminum in the UV region, the conductive reflective substrate consists of aluminum material. The upper aluminum metal surface is etted into a periodic plasmonic metamaterial grating structure, where $h_{Al}$ = 30 nm, $w_{Al}$ = 60 nm, and $p_{Al}$ = 180 nm. The graphene and plasmonic metamaterials are separated by air. The graphene ends are supported by an insulating layer, SiO$_2$, and a graphite dissociated surface lubrication layer. Due to the existence of the singular saddle point, the graphene suspended in the uppermost layer has an asymmetric light absorption peak near 4.6 eV (270 nm), which is close to 9\% \cite{yang2009excitonic,boosalis2012visible,cai2018ultraviolet,xu2020ultraviolet}, far higher than the absorption of 2.3\% \cite{jiang2018broad,liu2019two} in the visible and infrared regions, which greatly improves the optical response of graphene in the UV region. According to the imaginary part of the permittivity of graphene at 4.6 eV provided in Ref. \cite{yang2009excitonic}, the real part of the permittivity can be expressed in the integral form of the imaginary part of the permittivity by Kramers--Kronig (KK) relations \cite{lucarini2005kramers} $,
Re[\epsilon(\omega)]=1+\dfrac{1}{\pi}P\int_{0}^{\infty} \dfrac{s\ Im[\epsilon(\omega)]}{(s^2-\omega^2)}
ds$, where $P$ is the principal value integral. The permittivity of graphene in the UV region can be obtained as a function of frequency. The refractive index of graphene $n(\omega)=\sqrt{\epsilon(\omega)}$, and its thickness is 0.34 nm. Therefore, to better regulate the restoring force, the design of the lubrication layer on the graphite dissociation surface shown in Fig. \ref{Fig1} (a) is consistent with that in Ref. \cite{zhou2021ultrawide}. It consists of four L-shaped graphite dissociation surfaces. Similar air bridge structures have been successfully realized in previous experiments \cite{cartamil2018graphene,bao2012situ}. Thus, this structure is convenient for experimental verification.

When a bias is applied between the graphene and plasmonic metamaterials, the graphene and plasmonic metamaterials attract each other, causing them to experience a downward force. The surface is approximately parallel to the substrate because graphene bending is very small, and the size of the graphene is much larger than the distance between the graphene and plasmonic metamaterials. Thus, the structure can be approximated as a parallel plate capacitor. Furthermore, when a voltage is applied, the system suffers from a retraction force provided by interlayer binding energy, very little friction, and the attraction between graphene and plasmonic metamaterials. The equilibrium position of graphene can be fixed and calculated once these forces reach equilibrium conditions. The motion of graphene can be obtained by solving Newton's equations numerically, as described in Ref. \cite{zhou2021ultrawide}. The interlayer bonding energy of graphene used in the calculation is 0.23 J/m$^2$ \cite{liu2012interlayer}, and the friction coefficient is 0.006 MPa$\cdot$ s/m \cite{yang2013observation}.

The finite difference time-domain method is used to simulate the reflection characteristics of the structure. Periodic boundary conditions apply in $x$ and $y$ directions. Plane waves incident from the $z$ direction use the perfect matching layer to absorb all light outside the boundary along the propagation direction. Non-uniform grids are used to balance storage and computing time.

\section*{Results and Discussion}
\begin{figure}
	\centering
	\includegraphics[width=\linewidth,clip]{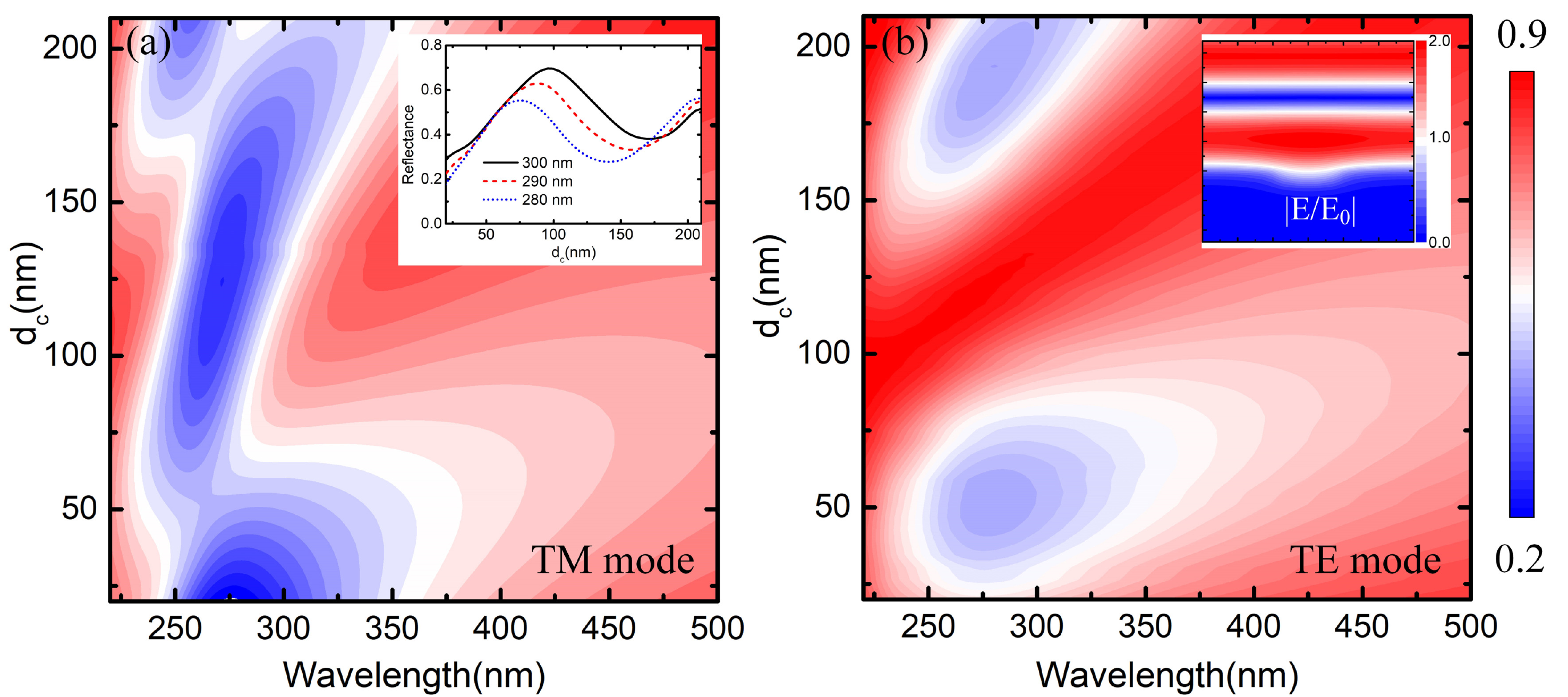}
	\caption{Reflection with the equilibrium position and wavelength of graphene for (a) TM mode and (b) TE mode; the inset of (a) shows the reflection changes with distance when the wavelength is 280, 290, and 300 nm, respectively; the inset of (b) shows electric field distribution.}
	\label{Fig2}
\end{figure}

The incident light can be divided into TM and TE modes according to their polarization. The TM (TE) mode is the magnetic (electric) component parallel to the metamaterial grating or graphene interface. When the TM-mode UV light incident from above the graphene along the $z$ direction, the diffraction phenomenon occurs on the surface of the plasmonic metamaterials, resulting in diffraction light waves. However, when the wave vector of the diffused light wave is consistent with that of the surface plasma wave, the surface plasma wave will be excited at the interface between the surface of the plasmonic metamaterial and the air medium. The phenomenon of surface plasmon resonance will be generated. Under the resonance excitation, the surface plasmons of plasmonic metamaterials can break through the diffraction limit and restrict the incident light to the sub-wavelength scale; thereby, forming a strong local field effect. Furthermore, when a voltage is applied between the plasmonic metamaterials and graphene, the graphene and plasmonic metamaterials are attracted to each other by Coulomb forces, which force them downward. Meanwhile, when no bias voltage is applied, graphene is leveled by restoring force. Thus, the graphene and plasmonic metamaterials attract or repel each other under the action of voltage; thereby, causing the graphene to slide on the graphite lubrication layer; the distance between the graphene and plasmonic metamaterials changes. The reflectance of the structure in TM mode varies with the distance and wavelength between graphene and plasmonic metamaterials (Fig. \ref{Fig2}(a)). Due to the local surface plasma of the metamaterial in this direction, the graphene moves from 20 to 100 nm away from the metamaterial surface at a wavelength of 300 nm, enabling efficient modulation only at 80 nm. This is because of the strong local field effect of light on the surface of the plasmonic metamaterials, which allows the higher modulation depth to be achieved only by moving the graphene a very small distance. However, when TE polarized light is incident, its field intensity direction is parallel to the interface between metamaterial grating and air. Thus, it will not influence the movement of free electrons on its surface. In other words, it will not produce a plasma resonance phenomenon at the interface. Fig. \ref{Fig2}(b) shows the variation of light reflectance with the distance and wavelength between graphene and plasmonic metamaterials under TE polarization. The illustration shows the electric field distribution, which is a standing wave distribution without local field generation. This demonstrates that the design of metamaterial grating makes the structure sensitive to polarization and makes it possible to develop new polarization related devices in UV region.

\begin{figure}
	\centering
	\includegraphics[width=8cm,clip]{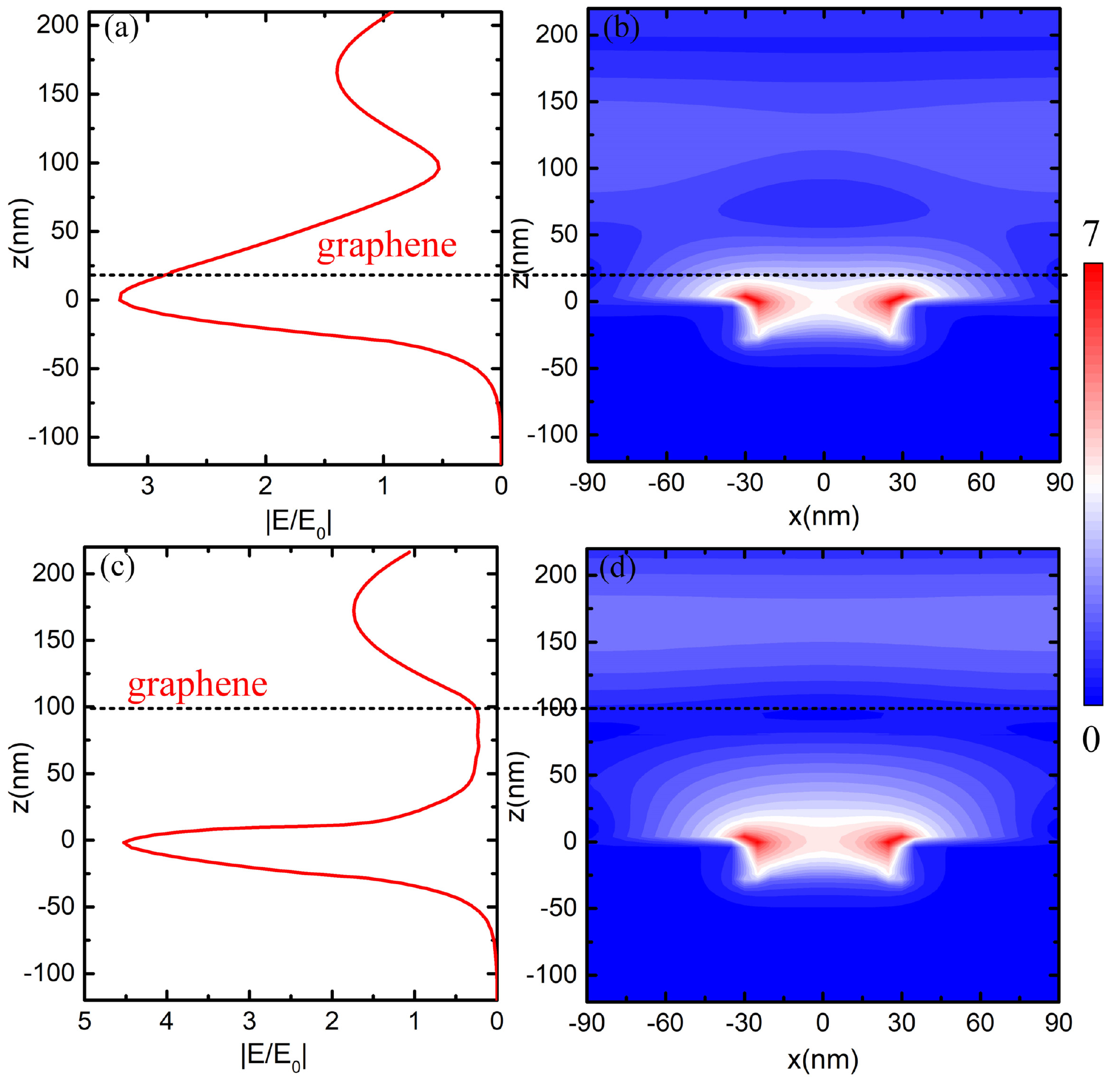}
	\caption{When $\lambda$ = 300 nm, (a) the electric field distribution of graphene equilibrium position at $z$ = 20 nm, (b) Electric field distribution at equilibrium position of graphene at $z$ = 100 nm.}
	\label{Fig3}
\end{figure}

Graphene absorption is closely related to the structure electric field distribution. Therefore, to analyze the physical mechanism behind the modulation depth of plasma resonance enhancement, we calculated the electric field distribution of graphene at 20 and 100 nm from the plasmonic metamaterials in TM mode. At a wavelength of 300 nm, when the equilibrium position of graphene is 20 nm above the plasmonic metamaterials, Fig. \ref{Fig3}(a) shows that local surface plasmon resonance is formed on the surface of the metal metamaterial structure, which usually occurs at the interface between the metal metamaterial structure and the medium. This strong localization will lead to the electric field enhancement effect and light energy convergence in the near field, which significantly enhances the interaction between light and matter. Moreover, this effect limits the electromagnetic field enhancement to the near field and extends to the nearby air medium, followed by rapid attenuation. Thus, there is no local field effect far away from the plasmonic metamaterials. Therefore, the gradient field with great difference in electric field distribution is formed. Graphene will show different absorption properties depending on the intensity of the electric field. Fig. \ref{Fig3}(b) shows that when graphene is located at 20 nm above the plasmonic metamaterials, i.e., when graphene is in a strong electric field, the absorption of graphene is large, resulting in a small reflection of the device. However, when graphene is far away from the plasmonic metamaterials, and its equilibrium position is about 100 nm above the plasmonic metamaterials, graphene is in a weak electric field with small absorption, so the light reflectivity of the device is high (Figs. \ref{Fig3}(c) and (d)). In this way, the distance between the graphene and plasmonic metamaterials can be changed by applying a voltage. The light absorption of the graphene can be adjusted by the electric field distribution gradient between them; thereby, modulating the reflectivity of the modulator and enabling efficient and fast light modulation.

\begin{figure}[htp]
	\centering
	\includegraphics[width=12cm,clip]{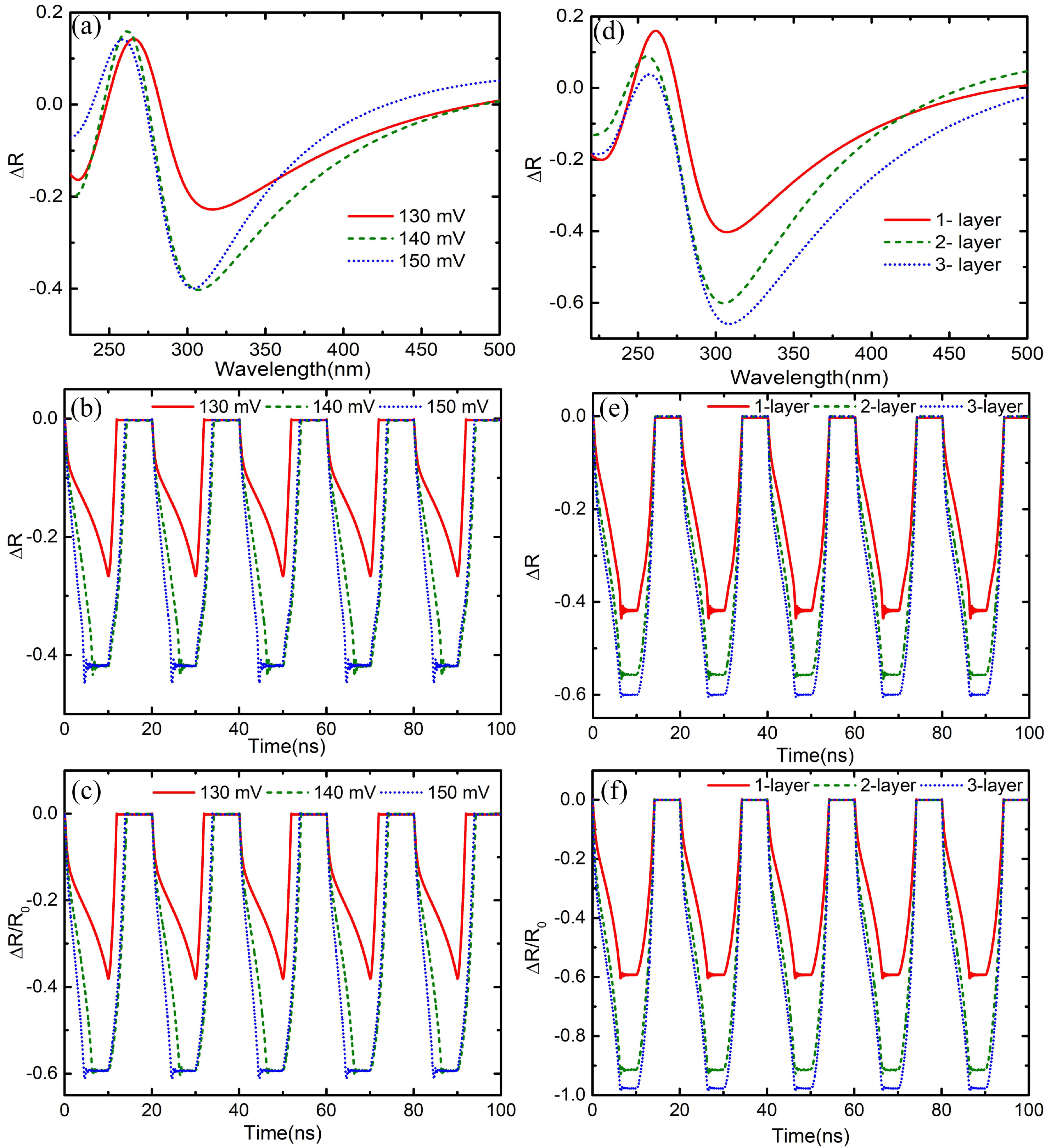}
	\caption{(a) Absolute variation of reflection with wavelength at different voltages. (b) The absolute change of reflection under different voltages varies with time at 5 ns. (c) The relative change of reflection with time under different voltages at 5 ns. (d) The absolute variation of reflection with wavelength at different graphene layers. (e) The absolute change in reflection of different graphene layers over time at 5 ns. (f) The relative change of the reflection of different graphene layers over time at 5 ns.}
	\label{Fig4}
\end{figure}

Fig. \ref{Fig4}(a) shows the absolute change of reflection with wavelength at different voltages. There is a peak near the wavelength of 300 nm, indicating that graphene has a large absorption change at this wavelength due to the strong local field of light formation in the plasmonic metamaterials. As the voltage decreases, the absolute change peak of reflection redshifts, but the modulation depth decreases gradually. Figs. \ref{Fig4}(b) and \ref{Fig4}(c) show the absolute and relative change of reflectivity when the driving electric field is a square wave with a period of 20 ns, respectively. $\Delta R = R-R_0$ is the change of reflectivity, and the relative change of reflectivity is $\Delta R/R_0$, where $R_0$ refers to the reflectivity under zero bias voltage. Here, the initial position of graphene is set to 100 nm away from the plasmonic metamaterials. The response time of the modulator is about 4 ns, which is more than four orders of magnitude higher than the traditional mechanical modulation speed. The absolute change of reflection is more than 0.4, and the relative change of reflection is about 0.6. Due to the high absorption of graphene in the UV region \cite{yang2009excitonic} and the local field effect of metamaterials, the modulation depth of the UV modulator can be increased by 8.5 times compared with the graphene-based modulator in the visible and infrared regions. Meanwhile, the modulation voltage is only 150 mV, about 1$\sim$3 orders of magnitude smaller than the traditional optical modulation voltage. When the voltage changes to 140 mV, the response time is slightly shortened, indicating good voltage adaptability. When the modulation voltage drops to 130 mV, the driving force of the electric field becomes weak; the motion amplitude of graphene becomes small, and the modulation speed and depth decrease. This means that a very low voltage will significantly reduce the deformation of graphene. The interaction between graphene and plasma metamaterials can be approximated as a parallel plate capacitor. The force between parallel plate capacitors is proportional to $U^2/h$, where $U$ is the voltage between the two plates, and $h$ is the distance between the plates. The deformation in this study is much smaller than that of the deformation-based graphene modulation system. The deformation energy consumption of the device reduces significantly because the deformation energy consumption is proportional to the square of the deformation. This is also the reason why mechanical modulation based on the superlubricity of graphene can significantly reduce the spontaneous oscillation of graphene compared to traditional modulation based on the deformation of graphene \cite{zhou2021ultrawide}.

Since graphene absorption is proportional to the number of graphene layers, multilayer graphene can further improve light modulation depth. For monolayers, the absolute change of reflection is about 0.4, and the relative change is about 0.6; in contrast, for bilayers (three graphene layers), the absolute change is about 0.5, (0.58) and the relative change is about 0.9 (0.98) (Figs. \ref{Fig4}(d)-(f)). The modulation depth of the device improves as the number of graphene layers increases. The modulation wavelength range of two-layer and three-layer graphene is the same as that of monolayer graphene (Fig. \ref{Fig4}(d)). As the number of graphene layers increases, the absolute and relative changes of device reflection gradually increase. In addition to increasing the number of graphene layers, the modulation depth can be further improved using graphene and other two-dimensional materials with large absorption, such as heterojunction structures formed with h-BN \cite{chen2020wafer}.

\begin{figure}[htp]
	\centering
	\includegraphics[width=8 cm,clip]{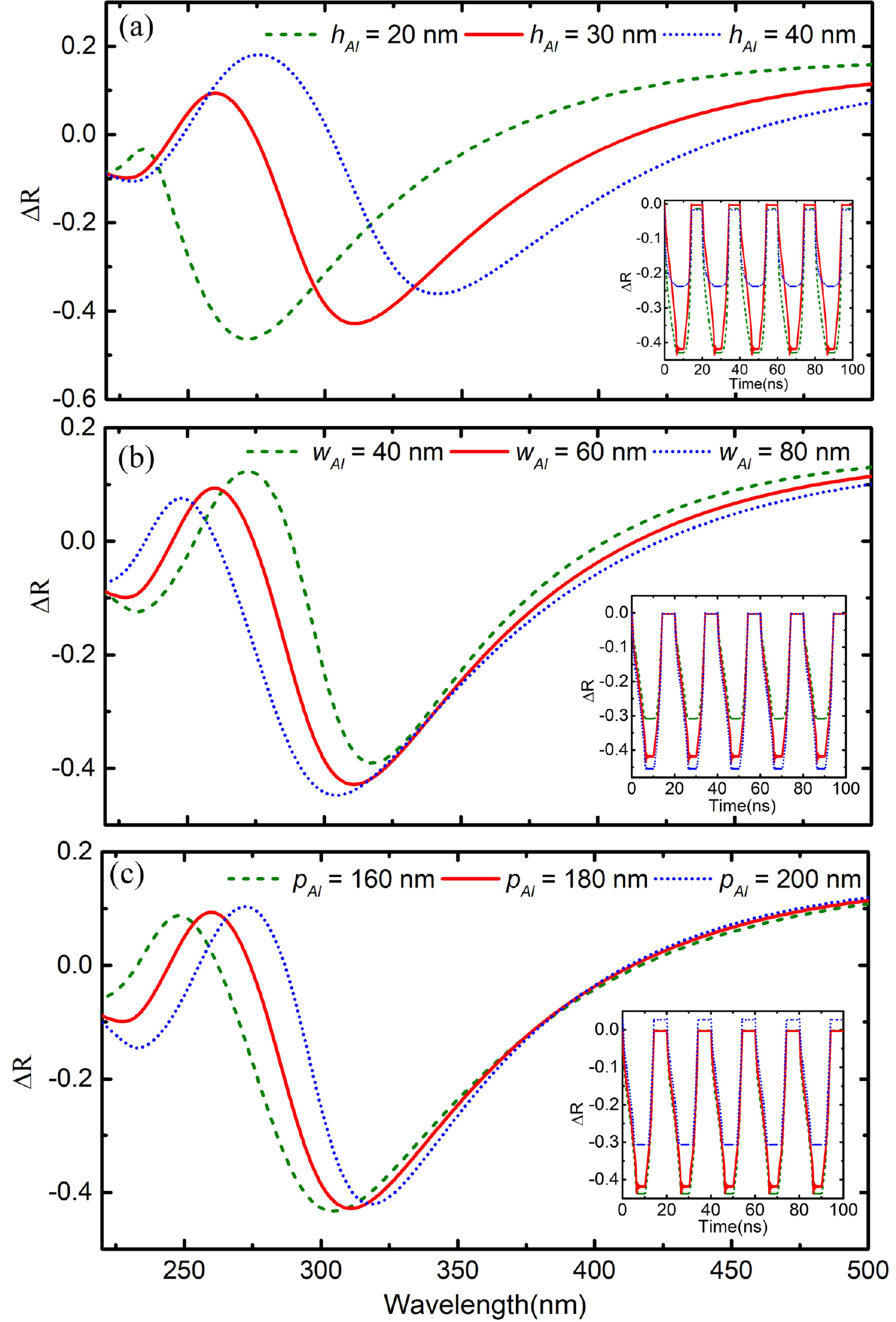}
	\caption{ (a) Absolute change of reflection varies with the height $h_{Al}$ of the plasmonic metamaterial grating. (b) The absolute change of reflection varies with the width $w_{Al}$ of plasmonic metamaterial grating. (c) The absolute change of reflection varies with the change of period $p_{Al}$ of plasmonic metamaterial grating. The illustration shows the absolute change of reflection over time with different parameters at a wavelength of 300 nm.}
	\label{Fig5}
\end{figure}

Furthermore, the local surface plasmon resonance shows adjustable optical properties according to the size of plasmonic metamaterial grating. Therefore, the influence of the structural parameters of plasmonic metamaterials on the device light modulation was investigated (Fig. \ref{Fig5}). The absolute change of the device's reflection varies with the depth of the plasma metamaterial grating (Fig. \ref{Fig5}(a)). With an increase in $h_{Al}$, the formant shifts redshift and changes sharply, confirming that the device is sensitive to the depth of the plasmonic metamaterial grating. Figs. \ref{Fig5}(b) and (c) show the influence of absolute change of reflection on the width of plasma metamaterial grating $w_{Al}$ and periodic $p_{Al}$. With an increase in $w_{Al}$, the peak value shifts blue; in contrast, the peak value redshifts as $p_{Al}$ increases. Since changing the parameters of the plasmonic metamaterial will affect the surface local field and graphene absorption, its modulation amplitude also changes accordingly, as shown in the illustration. Thus, the resonant wavelength and modulation depth can also be tuned by modulating the structural parameters of the plasma metamaterial.

\begin{figure}[htp]
	\centering
	\includegraphics[width=\linewidth,clip]{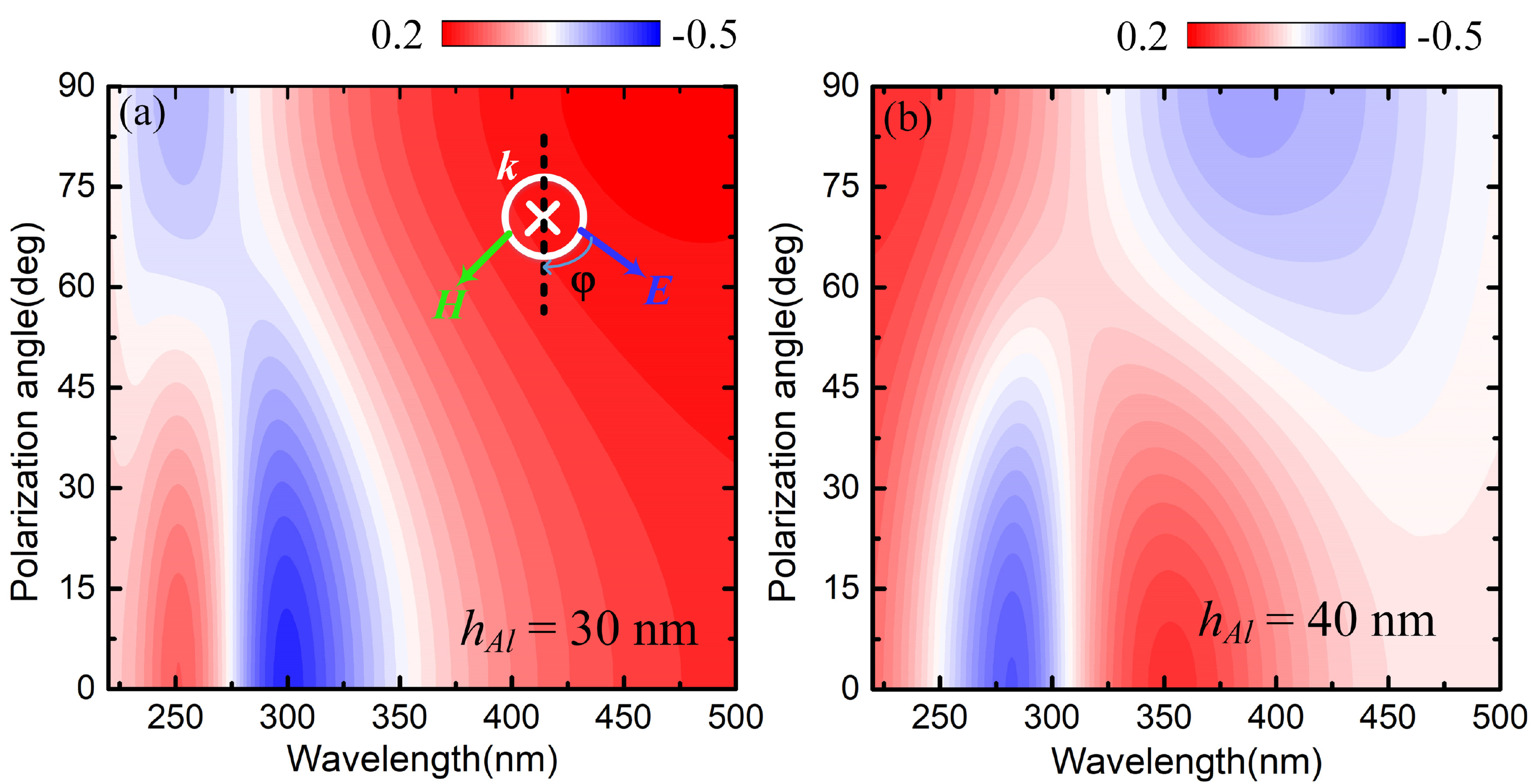}
	\caption{Absolute variation of reflection at different polarization angles: (a) $h_{Al}$ = 30 nm (b) $h_{Al}$ = 40 nm}
	\label{Fig6}
\end{figure}

To further investigate the polarization dependence of the modulator, we plot the spectra of the absolute change of reflection between graphene and plasmonic metamaterials at different polarization angles. As shown in Fig. \ref{Fig6}, this structure is sensitive to the polarization of incident light. When $h_{Al}$ = 30 nm, and the polarization angle (which is the included angle between the electric field and the $y$-axis) changes from $0^\circ$ to $90^\circ$, the absolute change of reflection decreases; the bandwidth decreases; the absolute change peak of reflection shifts blue. The absolute reflection changes the most when the wavelength is about 300 nm; its value changes from -0.44 to 0 in TM to TE modes, respectively. However, when $h_{Al}$ = 40 nm, the peak of absolute reflection change redshift (Fig. \ref{Fig6}(b)). A large change in device performance can be observed when $h_{Al}$ changes, with the absolute change of reflection varying from -0.22 in TM mode to 0 in TE mode at a wavelength of 300 nm. The wavelength of the maximum absolute change of reflection shifts to 280 nm, with the absolute change of reflection varying from -0.35 in TM mode to 0 in TE mode. According to the above analysis, the structure is more sensitive to $h_{Al}$ changes, which is attributed to the formation of the micro-cavity effect between plasmonic metamaterial and graphene. The device can also be tuned reflectively by polarization angle, which gives the structure flexible polarization tunability.

It is worth noting that although our results are obtained through simulation, all parameters used in the simulation are based on experimental results, such as friction coefficient and interlayer binding energy coefficient. Therefore, it is completely achievable experimentally. The design provides theoretical guidance for developing a UV modulator based on a graphene micromechanical system.

\section*{Conclusion}
In this study, we used the superlubricity of graphene and surface plasmonic metamaterial to realize UV electro-optic modulation of graphene NEMS. We used the strong localized light field formed by light near the plasmonic metamaterials to make the absorption of suspended graphene at different positions on the metamaterial surface be different. The graphene is periodically deflected up and down by periodic voltage drive. The numerical results show that the modulation depth of the electro-optic modulator can be significantly improved by combining the graphene in the UV region to form a surface plasmonic metamaterial structure with a strong local field. The absolute variation in reflection of monolayer graphene can be more than 0.4. The modulated reflection of three graphene layers has an absolute change of reflection of more than 0.6 and a relative change of reflection of 0.98. Additionally, the superlubricity further improves the performance of UV electro-optic modulator based on graphene NEMS due to its extremely low friction coefficient, with a modulation voltage lower than 200 mV and a modulation response speed of up to nanoseconds. The report indicates that graphene-plasmonic metamaterial devices with mechanical modulation and superlubricity will have important applications in UV communication fields.


%

\section*{Acknowledgements}This work was supported by the National Natural Science Foundation of China (NSFC) (Grant No. 62174040, 11764008), The 13th batch of outstanding young scientific and Technological Talents Project in Guizhou Province (No. [2021]5618), the Science and Technology Talent Support Project of the Department of Education in Guizhou Province (Grant No. KY[2018]045), the Young scientific and technological talents growth project of the Department of Education in Guizhou Province (Grant No. KY[2022]184), the Science and Technology Foundation of Guizhou Province, China (Grant No. [2020]1Y026), and Natural Science Research Project of Guizhou Minzu University (GZMUZK[2021]YB07).



\emph{Conflict of Interest}: The authors declare no competing
financial interest.

\end{document}